\documentclass{iaus}
\usepackage{graphicx}

\title[] %% give here short title %%
{3-D reconstructions of active stars -- observations}

\author[Korhonen]   %% give here short author list %%
{Heidi Korhonen$^{1, 2, 3}$}

\affiliation{
$^1$Niels Bohr Institute, University of Copenhagen, Juliane Maries
Vej 30, DK-2100 Copenhagen, Denmark\\[\affilskip]
$^2$Centre for Star and Planet Formation, Natural History Museum of Denmark, 
University of Copenhagen, {\O}ster Voldgade 5-7, DK-1350, Copenhagen, Denmark
\\[\affilskip]
$^3$Finnish Centre for Astronomy with ESO (FINCA), University of Turku, 
V{\"a}is{\"a}l{\"a}ntie 20, FI-21500 Piikki{\"o}, Finland
}

\pubyear{2013}
\volume{xxx}  %% insert here IAU Symposium No.
\pagerange{xxx-xxx}
\jname{Title of your IAU Symposium}
\editors{A.C. Editor, B.D. Editor \& C.E. Editor, eds.}
\begin{document}

\maketitle

\begin{abstract}
Stars are usually faint point sources and investigating their surfaces and 
interiors observationally is very demanding. Here I give a review on the 
state-of-the-art observing techniques and recent results on studying interiors 
and surface features of active stars.

\end{abstract}

\firstsection % if your document starts with a section,
\section{Introduction}

The only star we can study easily with any spatial resolution is the Sun, and 
now with the STEREO spacecrafts we can do it even in 3-D. For other stars we are 
usually restricted to using indirect methods for getting information of their 
surface structures. Here a short review on methods for studying stellar 
interiors and surfaces and the main recent results are presented.

\section{Interiors}

Kepler has not only revolutionised the exoplanet research, but also stellar 
astrophysics. The short, one minute, cadence data is uniquely suited for 
studying stellar oscillations and thus the interiors of stars (see, e.g., 
\cite[Gilliland \etal\ 2010]{Gilliland10}). Recent Kepler results, based on a 
sample of 500 solar-like stars, show that even though distribution of observed 
radii is similar to the one predicted by models of synthetic stellar 
populations in the Galaxy, the distribution of observed masses seems to deviate 
significantly from the predicted distribution 
(\cite[Chaplin \etal\ 2011a]{Chaplin1}). This result raises interesting 
questions on the star formation rate and initial mass function. Studying the 
same sample of solar-like stars further reveals that the number of stars 
showing oscillations decreases significantly with increasing levels of activity
(\cite[Chaplin \etal\ 2011b]{Chaplin2}). This on the other hand implies that 
the magnetic fields can inhibit the amplitudes of solar-like oscillations.

\section{Surfaces}

The strength of the dynamo-created stellar magnetic fields and the frequency of
the associated phenomena (e.g., starspots and flares) is strongly linked to the
rotation of the star, with rapid rotation enhancing the field generation (e.g.,
\cite[Pallavicini \etal\ 1981]{Pallavicini81}). Many active stars, rapid 
rotation usually induced either by youth or binarity, show large starspots 
which are detectable even using photometric observations from small 
ground-based telescopes (e.g., \cite[Kron 1947]{Kron47}). In some cases 
the monitoring has become almost nightly after the implementation of automatic 
photometric telescopes in the late 1980's (e.g., 
\cite[Rodon{\`o} \& Cutispoto 1992]{rodono}; 
\cite[Strassmeier \etal\ 1997]{Strassmeier}). These observations give crucial 
information on the stellar activity cycles (e.g., 
\cite[Ol{\'a}h \etal\ 2009]{Olah}) and long-term spot longitudes (e.g., 
\cite[Lehtinen \etal\ 2012]{Lehtinen}), i.e., 2-D information where one 
dimension is longitude and the other time. With observations using multiple 
filters crude spot latitudes can be obtained (e.g., 
\cite[Roettenbacher \etal\ 2011]{Roettenbacher}), making the reconstruction 3-D 
(longitude, latitude and time). Similarly, the continuous, high precision, 
space-based light-curves allow for recovering latitudinal and longitudinal 
information in addition to the temporal variations.

For obtaining more detailed stellar surface structures Doppler imaging 
techniques have to be used (e.g., \cite[Vogt, Penrod \& Hatzes 1987]{Vogt}), 
giving a 2-D snapshot of the surface temperature with a reasonable resolution 
(e.g., \cite[Korhonen \etal\ 2010]{Korhonen1}). Combining many observations 
from different epochs adds the third dimension, namely time (e.g., 
\cite[Skelly \etal\ 2010]{Skelly}; \cite[Hackman \etal\ 2012]{Hackman}). 
Doppler images can also be used for studying stellar surface rotation and its 
dependence on latitude (for a recent review see 
\cite[Korhonen 2012]{Korhonen2}). If spectral lines formed at different heights
in the stellar atmosphere are used in the reconstruction, also vertical 
structures in starspots can be mapped (\cite[Berdyugina 2011]{Berdyugina}), 
coming closer to what is possible in the solar case.

\section{Conclusions and outlook}

Obtaining multi-dimensional infromation on the active stars, which are just 
point sources, is challenging but not impossible. Observations give us detailed
information, both temporal and spatial, on occurrence of starspots. In the 
future long baseline infrared/optical interferometry can offer a direct check 
on some of the results. Furthermore, while the interferometric imaging is 
restricted to the brightest and nearest stars, the method does not require 
large projected rotational velocities, like Doppler imaging does. This will 
open up new targets that have not been approachable earlier, significantly 
widening the parameter space for activity studies.

~\\
{\bf Acknowledgments}
H.K. acknowledges the support from European Commission's Marie Curie IEF 
Programme, and an IAU travel grant to participate the General Assembly.

\end{document}